\documentclass[twocolumn]{svjour3}
\usepackage[utf8]{inputenc}

\usepackage{graphicx}
\usepackage{subfig}
\usepackage{balance}
\usepackage{booktabs}
\usepackage{amsmath}
\usepackage{amssymb}
\usepackage{hyperref}
\usepackage{algorithm}
\usepackage{algorithmic}
\usepackage{todonotes}

\usepackage{listings}
\begin{document}

\title{Semantic Trails of City Explorations:\\How Do We Live a City}

\author{Diego Monti \and
Enrico Palumbo \and
Giuseppe Rizzo \and
Rapha\"el Troncy \and
Thibault Ehrhart \and
Maurizio Morisio}

\institute{Diego Monti \at
Politecnico di Torino, Corso Duca degli Abruzzi 24, 10129 Turin, Italy\\
Tel.: +39-011-0907087\\
Fax: +39-011-0907099\\
\email{diego.monti@polito.it}
\and
Enrico Palumbo \at
Istituto Superiore Mario Boella, Via Pier Carlo Boggio 61, 10138 Turin, Italy\\
EURECOM, Sophia Antipolis, 450 Route des Chappes, 06410 Biot, France\\
Politecnico di Torino, Corso Duca degli Abruzzi 24, 10129 Turin, Italy\\
\email{enrico.palumbo@ismb.it}
\and
Giuseppe Rizzo \at
LINKS Foundation, Via Pier Carlo Boggio 61, 10138 Turin, Italy\\
\email{giuseppe.rizzo@linksfoundation.com}
\and
Rapha\"el Troncy \at
EURECOM, Sophia Antipolis, 450 Route des Chappes, 06410 Biot, France\\
\email{raphael.troncy@eurecom.fr}
\and
Thibault Ehrhart \at
EURECOM, Sophia Antipolis, 450 Route des Chappes, 06410 Biot, France\\
\email{thibault.ehrhart@eurecom.fr}
\and
Maurizio Morisio \at
Politecnico di Torino, Corso Duca degli Abruzzi 24, 10129 Turin, Italy\\
\email{maurizio.morisio@polito.it}
}

\date{Received: date / Accepted: date}

\maketitle

\begin{abstract}
The knowledge of city exploration trails of people is in short supply because of the complexity in defining meaningful trails representative of individual behaviours and in the access to actionable data. Existing datasets have only recorded isolated check-ins of activities featured by opaque venue types. In this paper, we fill the gaps in defining what is a semantic trail of city exploration and how it can be generated by integrating different data sources. Furthermore, we publicly release two datasets holding millions of semantic trails each and we discuss their most salient characteristics. We finally present an application using these datasets to build a recommender system meant to guide tourists while exploring a city.

\end{abstract}

\section{Introduction}
\label{sec:introduction}

Location-based social networks (LBSNs) allow users to share their position with friends, or even publicly, by performing a \emph{check-in} when they visit a certain venue or point-of-interest (POI). A POI can be defined as an entity that has a somewhat fixed and physical extension, like a landmark, a building, or a city.\footnote{As formalized by \url{https://schema.org/Place}.} A check-in is typically associated with many information of potential interest for researchers specialized in different domains, from urban mobility to recommender systems. For example, many LBSNs classify their POIs in consistent taxonomies, that assign an explicit semantic meaning to each check-in. Furthermore, each venue has a physical location, which can be represented by its geographical coordinates, and each check-in is performed at a specific point in time.

The contribution of this work is threefold: we formally define what is a set of temporally neighboring activities, which we called semantic trail of check-ins, and how to generate it, we propose a mapping between the venue categories available in Foursquare and the corresponding Schema.org terms, and we introduce the Semantic Trails Datasets (STDs) which are two different datasets of semantically annotated trails created starting from check-ins performed on the Foursquare social network. Differently from other datasets already available, we analyzed the check-ins at our disposal in order to group them into sequences of activities. Furthermore, we enriched the datasets by adding valuable semantic information, that is the Schema.org terms corresponding to the Foursquare category of the associated venues as well as the GeoNames and Wikidata entities representing the city in which the check-in was performed.

The remainder of this paper is structured as follows. In Section~\ref{sec:related-work}, we review related works, while, in Section~\ref{sec:semantic-trails}, we introduce the procedure used to generate the STDs. Then, we analyze the main characteristics of our datasets in Section~\ref{sec:analysis} and we present a possible use case in Section~\ref{sec:use-case}. Finally, we conclude and outline future works in Section~\ref{sec:conclusion}.

\section{Related Work}
\label{sec:related-work}

Different authors have analyzed user-created geographical data obtained from LBSNs. In the following, we distinguish among works related to data-driven studies (Section~\ref{sec:data-driven}), next POI recommendation (Section~\ref{sec:next-poi}), and check-in datasets (Section~\ref{sec:other-datasets}).

\subsection{Data-driven Studies}
\label{sec:data-driven}

Several works exploited LBSNs for a data-driven understanding of cities and characterizing the social behaviors related to urban mobility. Noulas et al.~\cite{Noulas2011} relied on a spectral clustering algorithm to create a semantic representation of city neighborhoods and to identify user communities that visit similar categories of places. Li et al.~\cite{Li2013} performed a statistical study with the aim of unraveling the correlations among venue categories and their popularity using a large check-ins dataset with $2.4$ million venues collected from different geographical regions. Preo\c{t}iuc-Pietro et al.~\cite{Preotiuc-Pietro2013} proposed to create a semantic representation of an urban area by relying on a bag of venue categories: they used such a representation to define a similarity measure between cities. More recently, Rizzo et al.~\cite{Rizzo2017} exploited density-based clustering techniques on a dataset containing venue categories to create high level summaries of the neighborhoods.

\subsection{Next POI Recommendation}
\label{sec:next-poi}

Other studies relied on LBSNs data to create algorithms capable of providing personalized recommendations of venues. A typical task addressed in literature is the prediction of the next POI in which a user is likely to be willing to go during the exploration of a city.

Different approaches have been considered to address this problem: for example, Cheng et al.~\cite{Cheng2013} proposed an extension of the matrix factorization method capable of considering the temporal relations in the check-in sequence, as well as the spatial constraints from the user. Ye et al.~\cite{Ye2013} introduced a framework based on a mixed hidden Markov model capable of first suggesting the most relevant venue categories and then selecting the actual suggested POIs given the estimated category distribution. Feng et al.~\cite{Feng2015} addressed the next \emph{new} POI recommendation problem using a personalized ranking metric embedding method. More recently, Palumbo et al.~\cite{Palumbo2017} proposed to recommend venue categories using recurrent neural networks, while S\'anchez et al.~\cite{Sanchez2018} exploited cross-domain techniques.

\subsection{Check-in Datasets}
\label{sec:other-datasets}

Some check-in datasets collected from LBSNs are already publicly available. The NYC Restaurant Rich Dataset~\cite{Yang2013} includes check-ins of restaurant venues in New York City only, as well as tip and tag data collected from Foursquare from October 2011 to February 2012. The NYC and Tokyo Check-in Dataset~\cite{Yang2015} contains check-ins performed in New York City and Tokyo collected from April 2012 to February 2013, together with their timestamp, GPS coordinates and venue category. The Global-Scale Check-in Dataset (GSCD)~\cite{Yang2016} includes long-term check-in data collected from April 2012 to September 2013 only considering the 415 most popular cities on Foursquare. All the previous datasets were created by Yang et al.~\cite{Yang2015,Yang2016} and they are publicly available on the Web.\footnote{\url{https://sites.google.com/site/yangdingqi/home/foursquare-dataset}}

However, none of them is focused on the analysis of temporal sequences of check-ins. In contrast, our approach for constructing semantic trails is similar to the one proposed by Parent et al.~\cite{Parent2013}, but instead of relying on raw GPS data, we consider check-ins from LBSNs and we semantically annotate them.

\section{Semantic Trails Datasets}
\label{sec:semantic-trails}

In this section, we detail the process that we followed for building the Semantic Trails Datasets (STDs) from the collections of check-ins at our disposal, that will be described in Section~\ref{sec:analysis}. The exploited algorithm is publicly available in our GitHub repository,\footnote{\url{https://github.com/D2KLab/semantic-trails}} while the resulting datasets have been published on figshare~\cite{figshare}.

\subsection{Dataset Generation}
\label{sec:generation}

We based our generation strategy on two initial collections of check-ins obtained from Foursquare, namely the GSCD and a second one created by the authors. Each of these sources consists in two different files serialized using a tabular format. The first one contains the check-ins collected from the platform, while the second one lists the venues involved and their details. More formally, each check-in associates a specific user with a certain venue and a timestamp, which represents the point in time when the check-in was performed.

\begin{definition}
Given the space of venues $\mathcal{V}$, the space of users $\mathcal{U}$, the space of timestamps $\mathcal{T}$, a check-in $\mathbf{c} \in \mathcal{C}$ is a tuple $\mathbf{c} = (\nu, \upsilon, \tau)$, where $\nu \in \mathcal{V}$ is the venue in which the user $\upsilon \in \mathcal{U}$ was located at the timestamp $\tau \in \mathcal{T}$.
\end{definition}

In contrast, a POI is characterized by a unique identifier, its geographical coordinates, and a category selected from the Foursquare taxonomy.\footnote{\url{https://developer.foursquare.com/docs/resources/categories}}

\begin{definition}
Given the space of categories $K$, a venue or point-of-interest (POI) $\nu \in \mathcal{V}$ is a tuple $\nu = (\varphi, \lambda, \kappa)$, where $\varphi$ is the latitude, $\lambda$ is the longitude, and $\kappa \in K$ is the associated category.
\end{definition}

In the following, we define a semantic trail as a list of consecutive check-ins created by the same user within a certain amount of time. This definition is similar to the one of semantic trajectories proposed by Parent et al.~\cite{Parent2013}, but it considers LBSNs instead of GPS data.

\begin{definition}
A semantic trail $\mathbf{s} \in \mathcal{S}$ is a temporally ordered list of check-ins $\langle \mathbf{c}_1,\allowbreak \mathbf{c}_2,\allowbreak \dotsc,\allowbreak \mathbf{c}_n \rangle$ created by a particular user $\upsilon \in \mathcal{U}$, i.e., for each $i$, $\mathbf{c}_i = (\nu_i, \upsilon, \tau_i)$ where $\tau_i < \tau_{i + 1} \land \nu_i \neq \nu_{i + 1}$.
\end{definition}

In order to construct the semantic trails from the initial datasets, we processed the check-ins and we analyzed their timestamps, for obtaining an unambiguous time representation that also includes the time zone. To this end, we exploited the \texttt{ciso8601} Python library.\footnote{\url{https://github.com/closeio/ciso8601}}

Then, we grouped the check-ins by user and we sorted them according to their timestamp. From such ordered lists of check-ins we constructed the semantic trails by assuming that two check-ins that are not distant in time more than eight hours belong to the same trail, similarly to what has been done in~\cite{Choudhury2010}.

In Algorithm~\ref{alg:generate}, we list the procedure for creating the set $\mathcal{S}$, given the set of users $\mathcal{U}$, the set of check-ins $\mathcal{C}$, and the time interval $\delta \tau = 8\ \mathrm{hours}$. Please note that some check-ins will not be included in any trail because they are too distant in time and, therefore, they will not be part of the STDs.

\begin{algorithm}
\caption{Generation of the set $\mathcal{S}$.}
\label{alg:generate}
\begin{algorithmic}[1]
\REQUIRE $\mathcal{U} \neq \{\emptyset\} \land \mathcal{C} \neq \{\emptyset\} \land \delta \tau \doteq \tau_i - \tau_j$
\STATE $\mathcal{S} \gets \{\emptyset\}$
\FORALL{$\upsilon \in \mathcal{U}$}
\STATE $\mathbf s \gets \varnothing$
\FORALL{$\mathbf c_i \in \mathcal{C}_{\upsilon} : \tau_{i - 1} < \tau_i \land i > 1$}
\IF{$\tau_i < \tau_{i - 1} + \delta \tau$}
\IF{$\mathbf s$ is $\varnothing$}
\STATE $\mathbf{s} \gets \langle \mathbf c_{i - 1} \rangle$
\ENDIF
\STATE $\mathbf{s} \gets \mathbf{s} + \langle \mathbf c_i \rangle$
\ELSE
\IF{\NOT $\mathbf s$ is $\varnothing$}
\STATE $\mathcal{S} \gets \mathcal{S} \cup \{\mathbf s\}$
\STATE $\mathbf{s} \gets \varnothing$
\ENDIF
\ENDIF
\ENDFOR
\ENDFOR
\RETURN $\mathcal{S}$
\end{algorithmic}
\end{algorithm}

In addition to this algorithm, we applied three different filters before constructing the trails in order to remove suspicious check-ins, that may have been spoofed with the help of automated software.

We first ignored the check-ins performed by a certain user in the same POI multiple times in a row and we only considered the last one, because such repetitions cannot result in meaningful semantic trails. Then, we discarded the check-ins performed by the same user in less than one minute, as it is unreasonable to visit a venue in such a short amount of time.

Finally, we filtered out the check-ins that require an unrealistic speed for moving from a certain venue to the next one. In particular, we removed consecutive check-ins that are associated with a speed greater than Mach~1 ($\sim 343$~m/s), as this value is higher than the normal cruise speed of an airplane. We computed the distance between two venues by applying the haversine formula to their geographical coordinates~\cite{Sinnott1984}. This approach is similar to the one followed in~\cite{Yang2016}.

\subsection{Semantic Enrichment}
\label{sec:enrichment}

In order to enrich the available datasets, we identified the city where each venue is probably located by performing the reverse geocoding of its coordinates. To this purpose, we used the \texttt{reverse\_geocoder} Python library\footnote{\url{https://github.com/thampiman/reverse-geocoder}} and the geographical coordinates of all the cities with a population greater than 500 people or seat of a fourth-order administrative division as reported by GeoNames. We also obtained the corresponding entities from Wikidata and we included their URIs, if available, in the STDs by matching the English city names and the geographical coordinates, when their distance was less than 10 km. We were able to find a correspondence for the $84\%$ of the cities available in GeoNames.

Furthermore, we manually mapped the categories listed in the Foursquare taxonomy with the Schema.org vocabulary. If a Foursquare category cannot be mapped with a leaf, then we mapped it with an ancestor. The mapping has involved three domain experts who performed a two-stage process: the first has involved two experts and it has elicited mappings and doubts, the second has involved the three experts whose the one excluded from the first stage acted as meta-reviewer, validating the mappings and resolving inconsistencies by answering to doubts. The resulting mapping is available in our GitHub repository.\footnote{\url{https://github.com/D2KLab/semantic-trails/blob/master/mapping.csv}} In the STDs, we included both the original Foursquare category and the associated Schema.org entity for each venue.

\subsection{Output Formats}
\label{sec:output}

The final result of the aforementioned process is available in two different file formats. The first one is a comma-separated values file containing the fields detailed in Table~\ref{tab:fields}. The second one is an equivalent RDF Turtle version of the dataset.

The Foursquare user identifier has been anonymized by replacing it with a number. On the other hand, the identifier of the venue corresponds to its Foursquare URI and, therefore, it can be used to retrieve additional information. For each check-in we also provide the category of the venue as available in the Foursquare taxonomy and the corresponding Schema.org term. The GeoNames identifier corresponds to the city in which the venue is located, while the country code refers to the country associated with that city. Finally, the timestamp is expressed in the ISO 8601 format and it has been approximated, for privacy reasons, to the minute.

As an example of the CSV format, we report two semantic trails in Listing~\ref{lst:csv-format}.

\begin{lstlisting}[caption={The first two trails obtained from the GSCD.},label={lst:csv-format}]
trail_id,user_id,venue_id,venue_category,venue_schema,venue_geonames,venue_wikidata,venue_city_name,venue_country,timestamp
1,1,4ec656207ee537da7d220f91,4bf58dd8d48988d162941735,schema:Place,geonames:5125734,wd:Q3449083,Malverne,US,2012-04-03T18:19:00-04:00
1,1,4e753db3c65bb91db4493d78,4bf58dd8d48988d116941735,schema:BarOrPub,geonames:5117891,wd:Q3452120,Franklin Square,US,2012-04-04T00:15:00-04:00
2,1,4cc36d0ad43ba143071c60f8,4bf58dd8d48988d101951735,schema:Store,geonames:5125734,wd:Q3449083,Malverne,US,2012-04-07T12:40:00-04:00
2,1,4e418ddb887740a51b5572d6,4bf58dd8d48988d134941735,schema:PerformingArtsTheater,geonames:5125734,wd:Q3449083,Malverne,US,2012-04-07T12:46:00-04:00
\end{lstlisting}

\begin{table}
\caption{The fields available in the STDs.}
\label{tab:fields}
\begin{tabular}{@{}ll@{}}
\toprule
Field & Description \\ \midrule
trail\_id & The numeric identifier of the trail \\
user\_id & The numeric identifier of the user \\
venue\_id & The Foursquare identifier of the venue \\
venue\_category & The Foursquare identifier of the category \\
venue\_schema & The Schema identifier of the category \\
venue\_geonames & The GeoNames identifier of the city \\
venue\_wikidata & The Wikidata identifier of the city \\
venue\_city\_name & The name of the city \\
venue\_country & The code of the country \\
timestamp & The timestamp of the check-in \\ \bottomrule
\end{tabular}
\end{table}

\section{Statistical Analysis}
\label{sec:analysis}

We generated the STDs starting from two different collections of check-ins obtained from the Foursquare platform. The first one is the Global-Scale Check-in Dataset (GSCD), created by the authors of~\cite{Yang2016} and publicly available on the Web. The second one is a similar but more recent set of check-ins realized by the authors of this work, originally collected in the context of~\cite{Palumbo2017}.

More in detail, we retrieved the check-ins performed by the users of the Foursquare Swarm\footnote{\url{https://www.swarmapp.com}} mobile application and publicly shared on Twitter from the Twitter API. Then, we collected additional information associated with the check-ins, like the venue in which it was performed and its geographical coordinates, thanks to the Foursquare API.

We report some statistics regarding these initial collections of check-ins in Table~\ref{tab:original}. The GSCD contains more check-ins, as it was collected for an higher number of days in a period of great popularity of LBSNs. On the other end, our dataset is being enriched with new check-ins continuously, therefore we envision future releases of the STDs based on a future snapshot of our collection of check-ins.

\begin{table}
\centering
\caption{The number of check-ins, venues, and users, the time interval and the period of collection for the two initial sets of check-ins.}
\label{tab:original}
\begin{tabular}{@{}lrr@{}}
\toprule
& \multicolumn{1}{c}{GSCD} & \multicolumn{1}{c}{Ours} \\ \midrule
Check-ins & 33,263,631 & 12,473,360 \\
Venues & 3,680,126 & 1,930,452 \\
Users & 266,909 & 424,730 \\
Time & 532 days & 382 days \\
Start & 2012-04 & 2017-10 \\
End & 2013-09 & 2018-10 \\ \bottomrule
\end{tabular}
\end{table}

We constructed two different versions of the STDs by applying the procedure described in Section~\ref{sec:semantic-trails} to these initial datasets. The two STD versions are named after the year in which the collection phase ended, that is 2013 for the GSCD and 2018 for the snapshot of our collection of check-ins.

We list several statistics regarding the STDs in Table~\ref{tab:std}. It is possible to observe that the number of initial check-ins available in the GSCD has been greatly reduced in STD~2013, while it has been only slightly decreased in STD~2018. This result is associated with the different collection protocols of the GSCD and our initial dataset. In fact, we decided to start removing misbehaving users directly during the collection phase, in order to limit the number of calls to the Foursquare API. In details, we discarded users that performed two check-ins in less than a minute for two times, because we identified this as a typical non-human behaviour~\cite{Palumbo2017}.

\begin{table}
\centering
\caption{The number of check-ins, trails, venues, users, and cities included in the two STD releases.}
\label{tab:std}
\begin{tabular}{@{}lrr@{}}
\toprule
& \multicolumn{1}{c}{STD 2013} & \multicolumn{1}{c}{STD 2018} \\ \midrule
Check-ins & 18,587,049 & 11,910,007 \\
Trails & 6,103,727 & 4,038,150 \\
Venues & 2,847,281 & 1,887,799 \\
Users & 256,339 & 399,292 \\
Cities & 10,152 & 52,011 \\ \bottomrule
\end{tabular}
\end{table}

The radically different number of cities involved in the semantic trails can also be explained by analyzing the collection protocols. The authors of the GSCD only considered densely populated areas, while we looked for check-ins without applying any geographical filter. The differences in the number of trails and venues are consistent with the size of the initial dataset.

In Table~\ref{tab:filters}, we detail the number of check-ins removed because of the different filters during the creation of the STDs. We observe a similar effect of the filters on the two datasets: for instance, the constrain on the repetition of a venue is always the most selective one. However, the number of invalid check-ins is extremely different, because of the various approaches exploited during the collection of the initial check-ins.

\begin{table}
\centering
\caption{The number of check-ins removed because of the different filters that we applied for creating the STDs. The total number of invalid check-ins is not equal to the sum of the different categories because the sets are not disjoint.}
\label{tab:filters}
\begin{tabular}{@{}lrrrr@{}}
\toprule
& \multicolumn{1}{c}{STD 2013} & \multicolumn{1}{c}{STD 2018} \\ \midrule
Venue & 2,381,182 & 275,359 \\
Time & 1,627,688 & 96,879 \\
Speed & 66,796 & 13,021 \\
Total & 3,963,133 & 366,491 \\ \bottomrule
\end{tabular}
\end{table}

\begin{figure}
\centering
\subfloat[STD 2013]{\includegraphics[width=.48\textwidth]{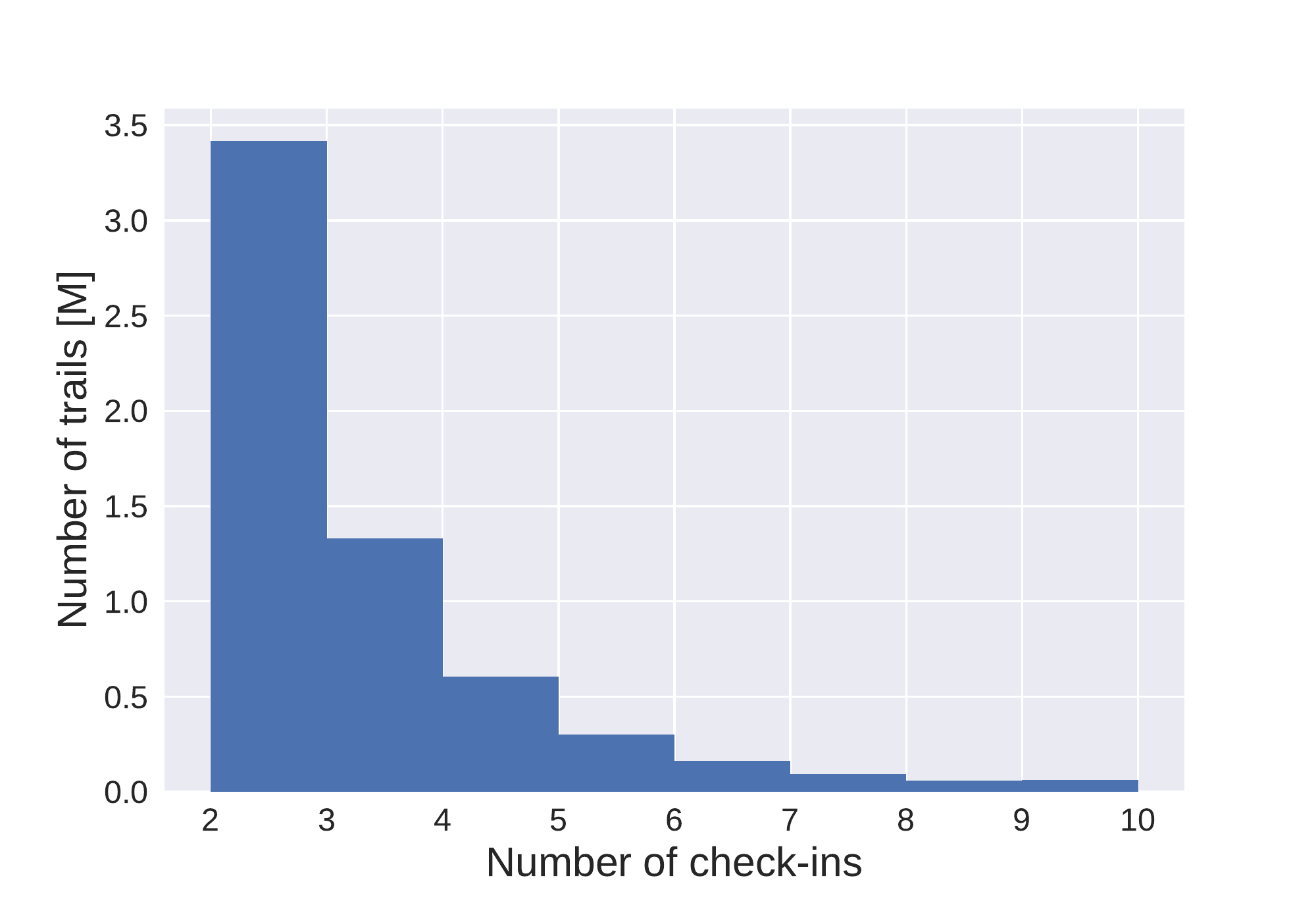}}\\
\subfloat[STD 2018]{\includegraphics[width=.48\textwidth]{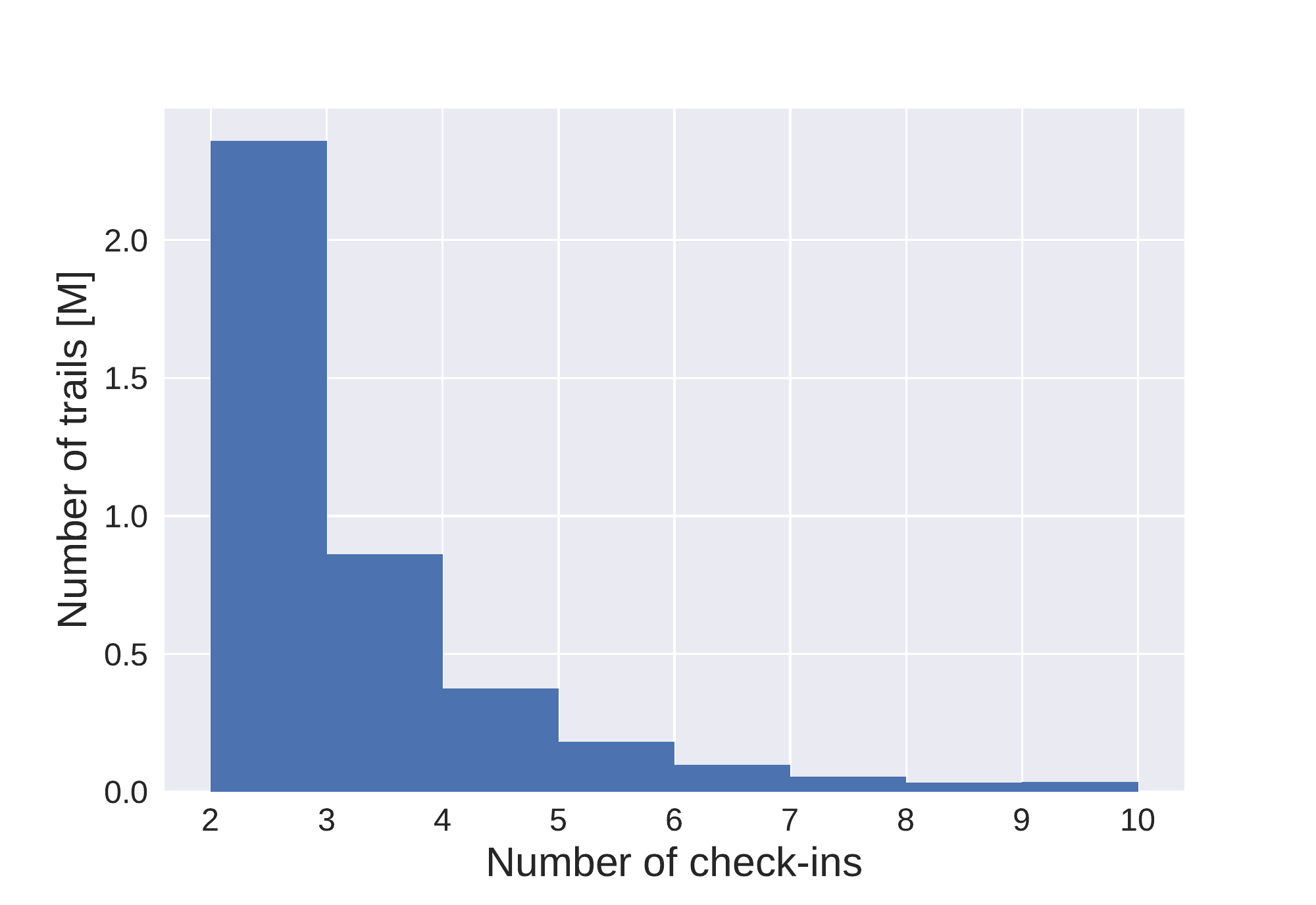}}
\caption{Histograms representing the distribution of trail lengths. We only considered trails with less than 10 venues due to graphical constraints. The scale of the vertical axis is in millions.}
\label{fig:lengths}
\end{figure}

Furthermore, we analyzed the lengths of the semantic trails that we built: the truncated histograms of their distributions are available in Figure~\ref{fig:lengths}. We observe that the distributions of the two datasets are similar, even if STD~2013 includes a higher number of trails. The average trail lengths are $3.05$ in STD~2013 and $2.95$ in STD~2018, while their standard deviations are $2.16$ and $1.99$ respectively.

\begin{figure}
\centering
\subfloat[STD 2013]{\includegraphics[width=.48\textwidth]{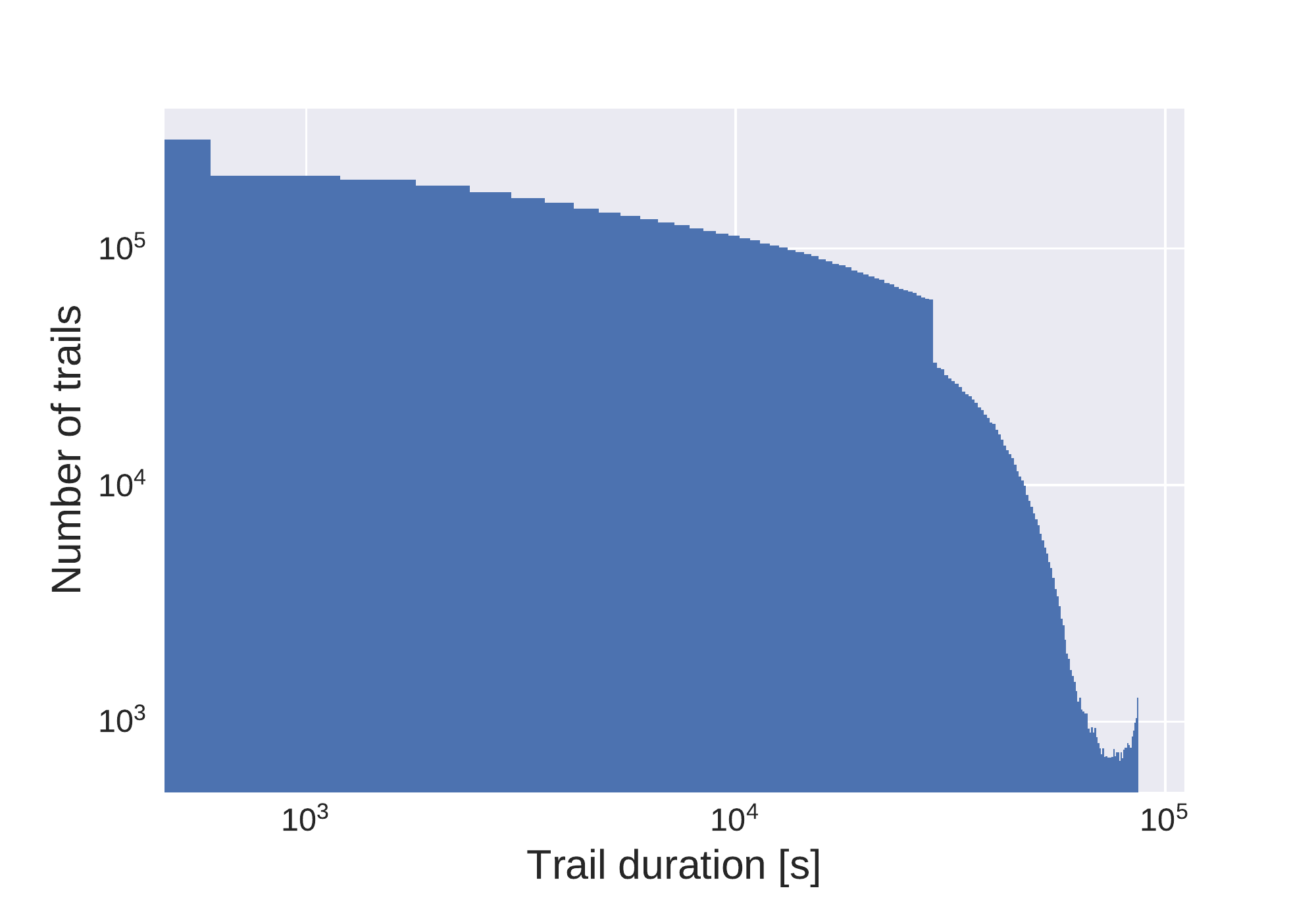}}\\
\subfloat[STD 2018]{\includegraphics[width=.48\textwidth]{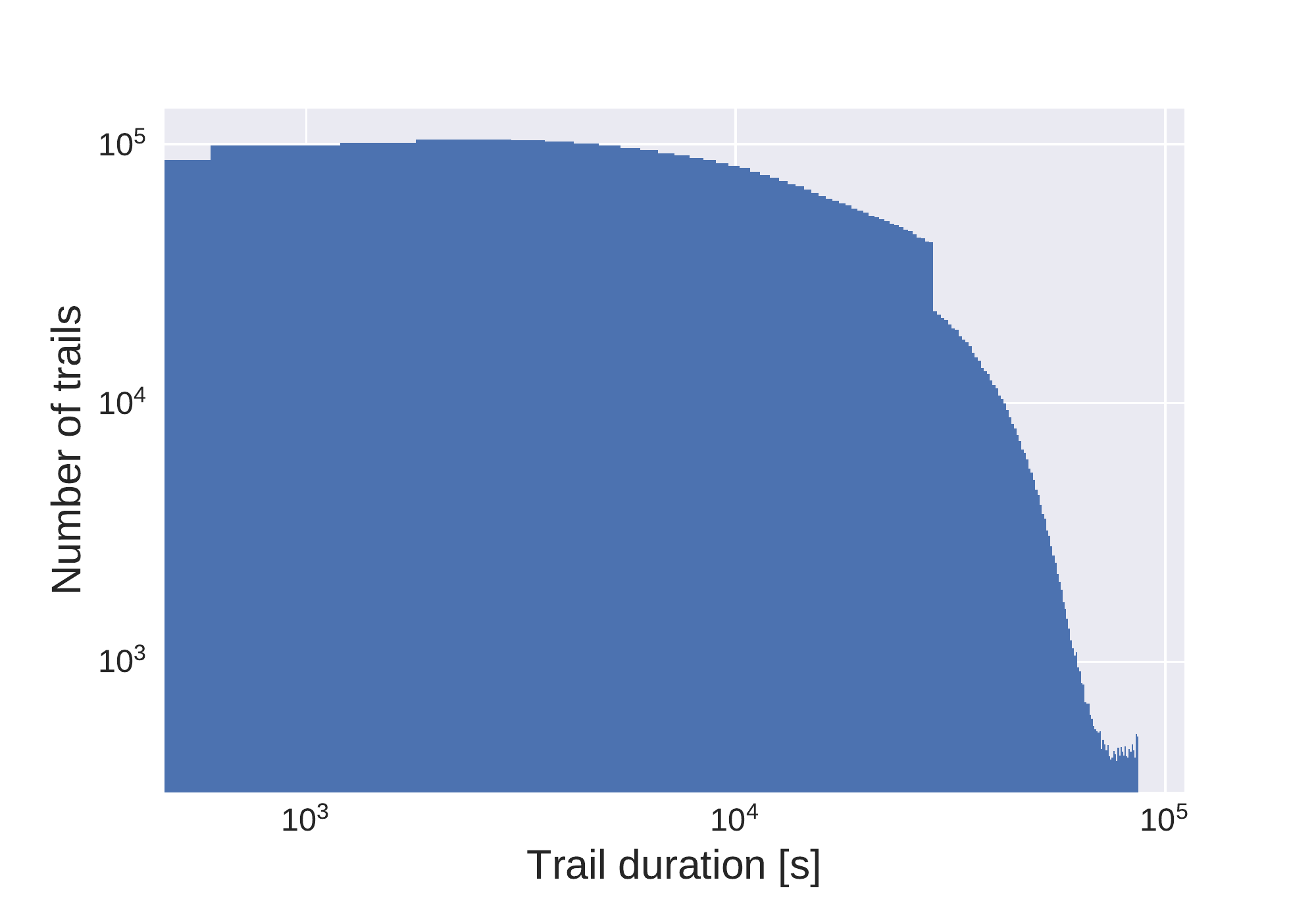}}
\caption{Histograms representing the distribution of trail time duration. We only considered trails lasting less than 24 hours due to graphical constraints. Both axes are represented in a logarithmic scale. The unit of the $x$ axis is seconds. The discontinuity of the curves is caused by the time limit used to build the trails.}
\label{fig:duration}
\end{figure}

We also depicted, in Figure~\ref{fig:duration}, the histograms representing the distributions of time durations, that is the number of time units between the first and the last check-in of a trail. It is interesting to notice that STD~2013 has a higher number of short trails, while STD~2018 contains more trails that have a relatively longer time duration with respect to very short ones. This difference may be explained by the fact that the platform and the behaviour of its users evolved during the years: longtime users may be more willing to share check-ins in a constant way.

\begin{figure}
\centering
\subfloat[STD 2013]{\includegraphics[width=.48\textwidth]{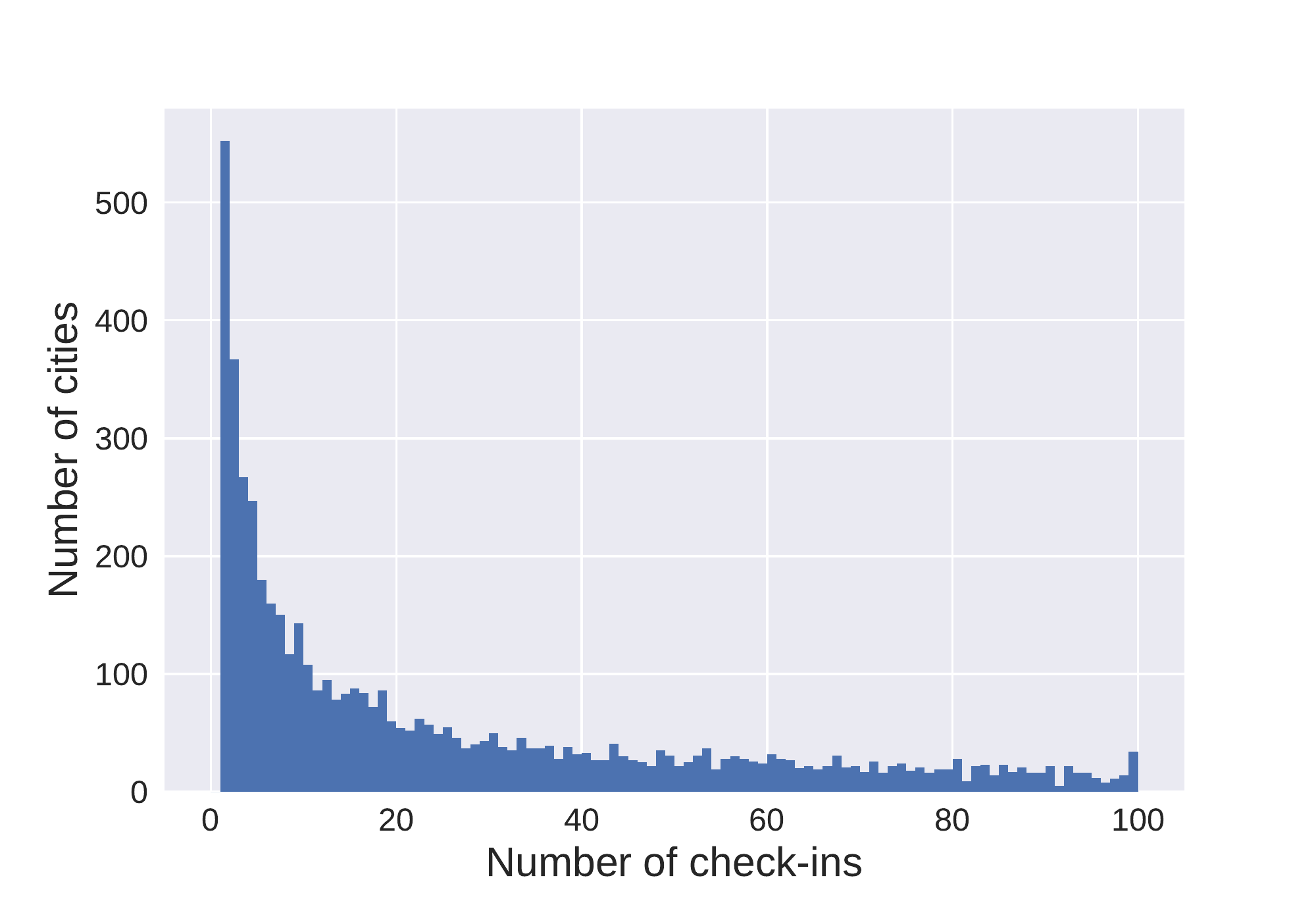}}\\
\subfloat[STD 2018]{\includegraphics[width=.48\textwidth]{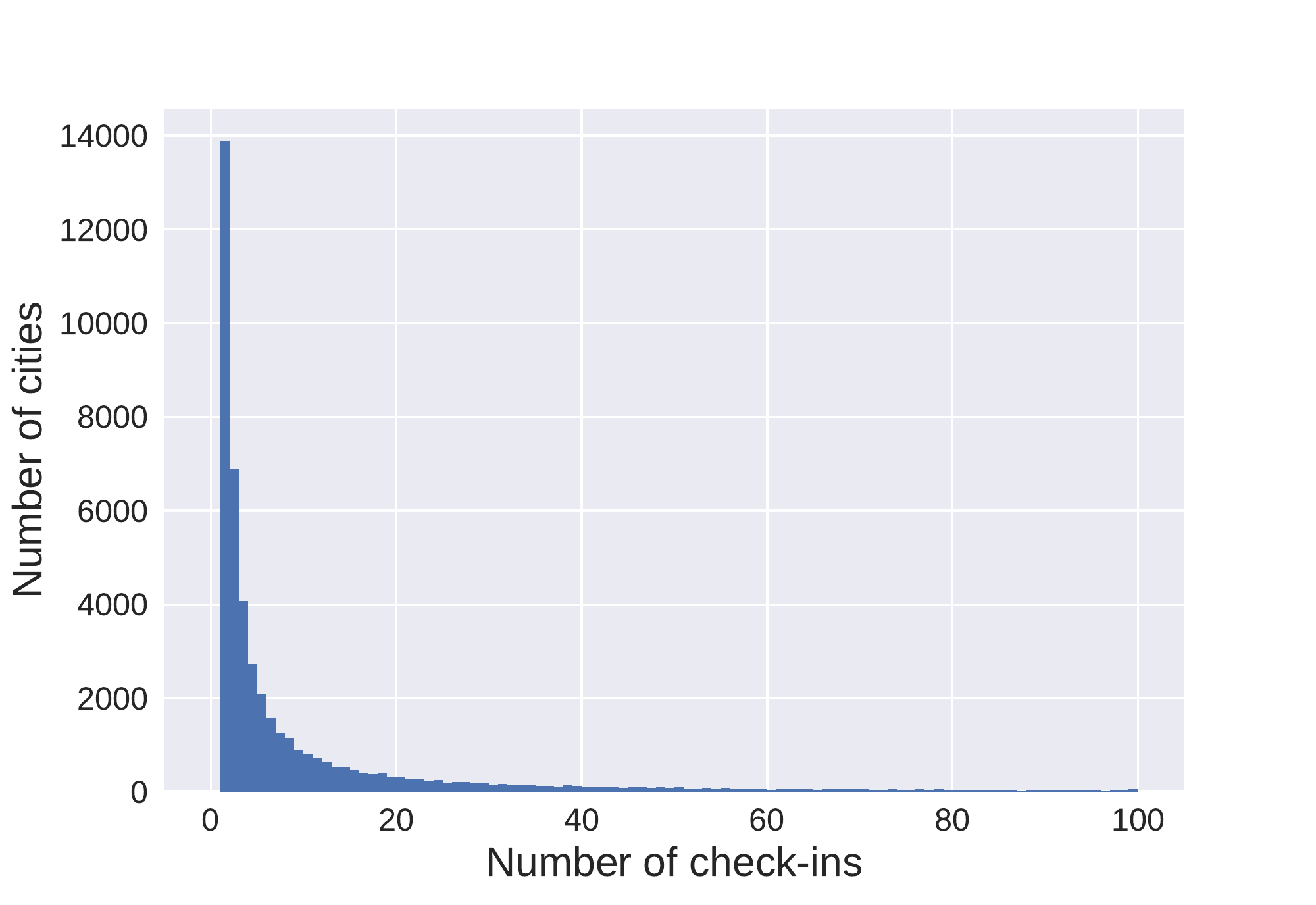}}
\caption{Histograms representing the distribution of cities per number of check-ins. We only considered cities with less than a hundred check-ins due to graphical constraints. The second dataset is more geographically widespread than the first one, as it contains an higher number of cities with a lower number of check-ins.}
\label{fig:cities}
\end{figure}

In order to analyze the check-ins of the two datasets from a spatial point of view, we considered the distributions of the number of check-ins for each city. As can be deduced from Figure~\ref{fig:cities}, STD~2013 includes a lower number of cities with less than a hundred check-ins, while STD~2018 contains many cities with a relatively low number of check-ins. This result is also related to the different number of cities available in the datasets as consequence of the initial collection protocol. For these reasons, STD~2018 may be more useful to characterize globally widespread behaviours, while the focus of STD~2013 is only on densely populated areas.

We also computed the number of check-ins for each country in the two STDs, which are reported in Table~\ref{tab:countries}. Some interesting differences emerge from these results: for example, Japan moved from the fifth to the first place in STD~2018, while Brazil was superseded by Malaysia. These observations can be easily explained by considering the different collection protocols and the possible changes in the usage patterns of the Foursquare platform during the years.

\begin{table}
\centering
\caption{The five countries with the highest number of check-ins in the two versions of the dataset.}
\label{tab:countries}
\subfloat[STD 2013]{\begin{tabular}{@{}lr@{}}
\toprule
\multicolumn{1}{l}{Country} & \multicolumn{1}{c}{Check-ins} \\ \midrule
Turkey & 3,282,073 \\
Brazil & 1,994,148 \\
USA & 1,859,310 \\
Malaysia & 1,584,552 \\
Japan & 1,553,603 \\ \bottomrule
\end{tabular}}\\
\subfloat[STD 2018]{\begin{tabular}{@{}lr@{}}
\toprule
\multicolumn{1}{l}{Country} & \multicolumn{1}{c}{Check-ins} \\ \midrule
Japan & 5,075,916 \\
Turkey & 2,030,934 \\
Kuwait & 801,867 \\
Malaysia & 764,733 \\
USA & 583,445 \\ \bottomrule
\end{tabular}}
\end{table}

Furthermore, we investigated the number of check-ins from STD~2018 in the two most popular countries, namely Japan and Turkey, grouped by the Schema.org category of their venue. The purpose of this analysis, whose results are listed in Table~\ref{tab:venues}, is to propose a simple but effective way of characterizing the different human behaviours that are typically associated with a certain culture. From these figures it is possible to observe that check-ins performed in train stations are very common in Japan, while in Turkey the most widespread category of venues is coffee shop.

\begin{table}
\centering
\caption{The five Schema.org venue categories with the highest number of check-ins from STD~2018 in Japan and in Turkey.}
\label{tab:venues}
\subfloat[Japan]{\begin{tabular}{@{}lr@{}}
\toprule
\multicolumn{1}{c}{Entity} & Check-ins \\ \midrule
schema:TrainStation & 1,198,732 \\
schema:Restaurant & 704,782 \\
schema:CivicStructure & 428,321 \\
schema:ConvenienceStore & 152,014 \\
schema:SubwayStation & 147,659 \\ \bottomrule
\end{tabular}}\\
\subfloat[Turkey]{\begin{tabular}{@{}lr@{}}
\toprule
\multicolumn{1}{c}{Entity} & Check-ins \\ \midrule
schema:CafeOrCoffeeShop & 371,118 \\
schema:CivicStructure & 179,997 \\
schema:Restaurant & 167,832 \\
schema:AdministrativeArea & 152,237 \\
schema:FoodEstablishment & 132,183 \\ \bottomrule
\end{tabular}}
\end{table}

In order to demonstrate the usefulness of a semantically annotated dataset, we computed additional statistics by also relying on external information obtained from GeoNames. In detail, we downloaded the number of inhabitants of the cities in which the check-is were performed, if available, and we considered the check-ins of small cities separately from the ones of big cities. We define a big city as a city with more than a hundred thousand inhabitants.

In Table~\ref{tab:std-cities}, we list the number of trails and check-ins in the STDs performed in small and big cities, while in Table~\ref{tab:venues-cities} we report the most frequent venue categories in STD~2018 grouped by the size of the cities. It is interesting to notice that airports are associated with small cities, as they are usually located outside densely populated areas.

\begin{table}
\centering
\caption{The number of trails and check-ins performed in small and big cities.}
\label{tab:std-cities}
\subfloat[Small cities]{\begin{tabular}{@{}lrr@{}}
\toprule
 & \multicolumn{1}{c}{Trails} & \multicolumn{1}{c}{Check-ins}\\ \midrule
STD 2013 & 2,045,440 & 4,444,930 \\
STD 2018 & 1,705,937 & 3,584,304 \\ \bottomrule
\end{tabular}}\\
\subfloat[Big cities]{\begin{tabular}{@{}lrr@{}}
\toprule
 & \multicolumn{1}{c}{Trails} & \multicolumn{1}{c}{Check-ins} \\ \midrule
STD 2013 & 4,693,791 & 12,350,172 \\
STD 2018 & 2,855,417 &  6,971,715 \\ \bottomrule
\end{tabular}}
\end{table}

\begin{table}
\centering
\caption{The five Schema.org venue categories associated with the highest number of check-ins performed in small and big cities in STD~2018.}
\label{tab:venues-cities}
\subfloat[Small cities]{\begin{tabular}{@{}lr@{}}
\toprule
\multicolumn{1}{c}{Entity} & Check-ins \\ \midrule
schema:CivicStructure & 433,467 \\
schema:Restaurant & 372,225 \\
schema:CafeOrCoffeeShop & 303,209 \\
schema:FoodEstablishment & 184,563 \\
schema:Airport & 171,464 \\ \bottomrule
\end{tabular}}\\
\subfloat[Big cities]{\begin{tabular}{@{}lr@{}}
\toprule
\multicolumn{1}{c}{Entity} & Check-ins \\ \midrule
schema:TrainStation & 1,005,890 \\
schema:Restaurant & 942,711 \\
schema:CivicStructure & 497,117 \\
schema:CafeOrCoffeeShop & 443,195 \\
schema:ShoppingCenter & 319,543 \\ \bottomrule
\end{tabular}}
\end{table}

\section{Tourist Sequence Recommender}
\label{sec:use-case}

The rich set of metadata collected in the STDs provides an explicit semantic meaning to users' activities. In fact, venue categories play an important role in POI recommender systems, as they enable to model user interests and personalize the recommendations~\cite{Liu2013}. The concept of trail, as defined previously, exploits the concept of temporal correlation that is a cornerstone for generating sequences of activities. In the past years, little attention has been dedicated to the temporal correlations among venue categories in the exploration of a city, which is nonetheless a crucial factor in recommending POIs. 

Take the example of a check-in in an Irish Pub at 8 PM: is the user more likely to continue the evening in a Karaoke Bar or in an Opera House? Better a Chinese Restaurant or an Italian Restaurant for dinner after a City Park in the morning and a History Museum in the afternoon? Note that generating these sequences requires an implicit modeling of at least two dimensions: temporal, as certain types of venues are more temporally related than others (e.g. after an Irish Pub, people are more likely to go to Karaoke than to a History Museum), and personal, as venue categories implicitly define a user profile, independently from their order (e.g. Steakhouse and Vegetarian Restaurant do not go frequently together). Most of existing studies attempt to model directly sequences of POIs rather than their categories to recommend the next POI to a user.

\begin{figure*}
\centering
\includegraphics[width=\textwidth]{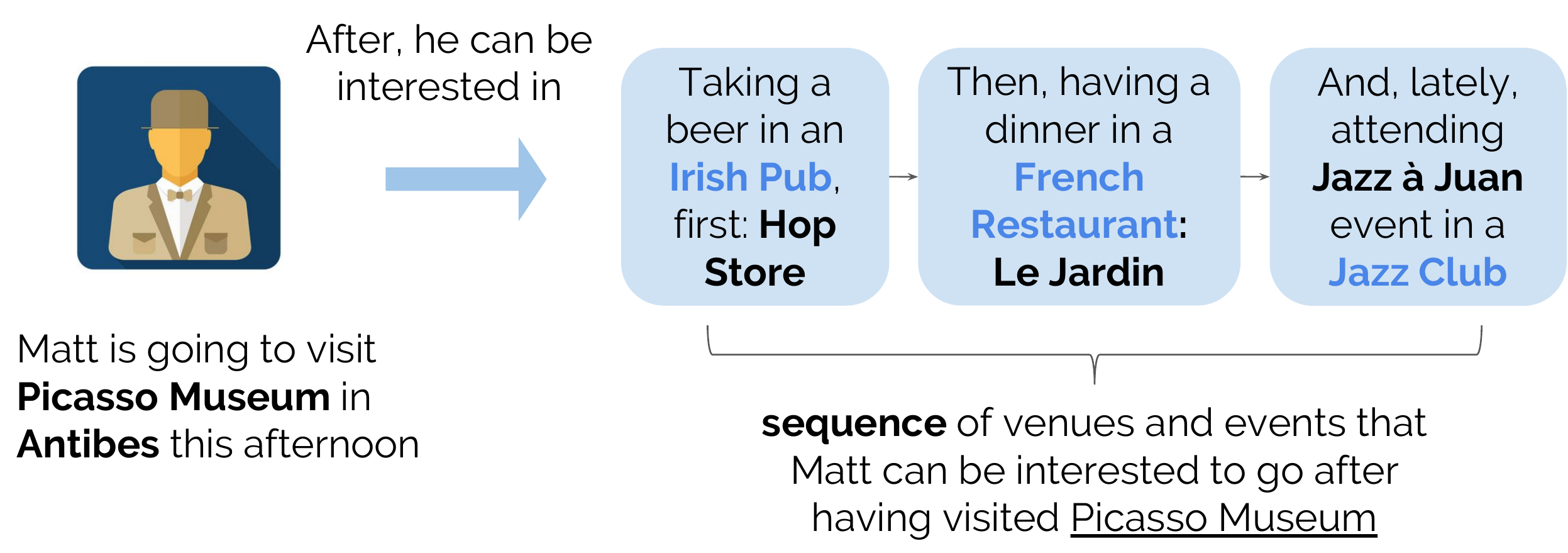}
\caption{The illustration of the Tourist Sequence Recommender in action: it takes as input a seed and it generates a sequence composed of places and events contextualized according to the city where the tourist is located.}
\label{fig:usecase}
\end{figure*}

In~\cite{Palumbo2017}, we presented an approach based on a neural learning model, and more precisely, Recurrent Neural Networks (RNNs), to generate sequences of tourist activities. The RNNs are trained with the sequential data available in the STDs and the output is expected to be a sequence of categories. The space of possible categories is defined by the Foursquare taxonomy, which classifies venues in a hierarchical taxonomy. In order to initiate the generation process, the neural learning model takes as input a seed, i.e. a category from which the tourist wishes to start his city exploration. Figure~\ref{fig:usecase} illustrates the process of semantic city trail generation. As it can be observed in the figure, the instantiating of places or events (entities) was considered as an integral part of the process and it was issued by querying the 3cixty knowledge base~\cite{Troncy2017}.

The impact of such use case was certified by a controlled and online experimentation with real users and it proved how impactful the STDs are in terms of meaningful resources to learn a model to generate tourist activity sequences and quality of metadata used to train our neural learning models.  

\section{Conclusion and Future Work}
\label{sec:conclusion}

In this work, we introduced the STDs, two datasets containing millions of check-ins performed on Foursquare and grouped into semantic trails, that are sequences of temporally neighboring activities. We described the algorithm used to generate such trails and we detailed the process followed to enrich the available data. We associated each check-in with the Schema.org term representing the venue category in which it was performed and we also identified the Wikidata entities corresponding to the city and country of the venue. We characterized the two datasets by analyzing them considering different dimensions and we demonstrated the usefulness of semantically annotated data by relying on external information to compute additional statistics. Finally, we briefly described a possible use case of such datasets, in which we proposed a tourist recommender system trained using the trails available in STD~2018. However, we envision different possible scenarios that could benefit from such datasets, for example human behaviour analysis and urban mobility studies.

The generation phase brought to further attention three points, namely the complexity of the mapping between Foursquare categories and Schema.org, the difficulties in obtaining a comprehensive list of cities, and the possible issues caused by inconsistencies present on Wikidata. We observed that different venue categories are not available on Schema.org: for this reason, they have been associated with the most similar term or with a common ancestor. Furthermore, even if some categories are available, they are not considered as a more specific type of schema:Place or schema:Event, and, therefore, they have been mapped with a general term. For example, schema:CollegeOrUniversity is considered an organization and not a physical place, so universities have been mapped with schema:CivicStructure. However, the mapping available in our GitHub repository is not meant to be final, and other researchers are invited to submit pull requests to improve it for future releases of the STDs. We also observed that there is no widespread entity that represents the concept of ``city'' on Wikidata. For this reason, we decided to rely on the definition provided by GeoNames, even if it considers some districts and neighborhoods as cities. We initially tried to rely on the DBpedia type dbo:City, but we empirically observed an high number of wrong or missing entities. Finally, we are aware of the fact that some URIs representing a city may be erroneous, due to duplicates or incorrect mappings between Wikidata and GeoNames. However, these problems can be fixed by future releases of our datasets if they are first resolved in the exploited knowledge base. These points will be part of future research activities.

\balance
\bibliographystyle{spmpsci}
\bibliography{references}

\end{document}